\documentclass[twocolumn,A4]{article}
\usepackage[dvips]{graphics}

\usepackage{setspace}
\usepackage{color}
\definecolor{gold}{rgb}{0.85,0.66,0}
\definecolor{dblue}{rgb}{0,0,0.8}

\begin{document}

\onecolumn

\begin{center}
{\bf{\Large {\textcolor{gold}{Multi-terminal Electron Transport Through 
Single Phenalenyl Molecule: A Theoretical Study}}}}\\
~\\
{\textcolor{dblue}{Paramita Dutta}}$^{\dag}$, {\textcolor{dblue}{Santanu 
K. Maiti}}$^{\dag,\ddag,}$\footnote{{\bf
Corresponding Author}: Santanu K. Maiti \\
$~$\hspace {0.45cm} Electronic mail: santanu.maiti@saha.ac.in} and 
{\textcolor{dblue}{S. N. Karmakar}}$^{\dag}$ \\
~\\
{\em $^{\dag}$Theoretical Condensed Matter Physics Division,
Saha Institute of Nuclear Physics, \\
1/AF, Bidhannagar, Kolkata-700 064, India \\
$^{\ddag}$Department of Physics, Narasinha Dutt College,
129 Belilious Road, Howrah-711 101, India} \\
~\\
{\bf Abstract}
\end{center}
We do parametric calculations to elucidate multi-terminal electron 
transport properties through a molecular system where a single phenalenyl 
molecule is attached to semi-infinite one-dimensional metallic leads. A 
formalism based on the Green's function technique is used for the 
calculations while the model is described by tight-binding Hamiltonian.
We explore the transport properties in terms of conductance, reflection 
probability as well as current-voltage characteristic. The most 
significant feature we articulate is that all these characteristics are 
very sensitive to the locations where the leads are connected and also 
the molecule-to-lead coupling strengths. The presence of other leads 
also has a remarkable effect on these transport properties. We study 
these phenomena for two-, three- and four-terminal molecular systems. 
Our numerical study may be utilized in designing tailor-made molecular 
electronic devices.

\vskip 1cm
\begin{flushleft}
\end{flushleft}

\newpage
\twocolumn

\section{Introduction}

Speed of growth of molecular electronics is being accelerated more 
and more as it has brought together scientists and engineers from 
various disciplines. The reason behind this attraction is inscribed into 
its smallness of size with wonderful electronic properties. In addition, 
several other properties such as magnetic, optical, etc., have been 
recognized in different molecules, which may be utilized in artificially 
tailored devices that are not possible with conventional 
materials~\cite{tao}. The concept of electron transport which emerged 
first in the theoretical work of Aviram and Ratner in $1974$~\cite{aviram} 
has opened a new era in the field of nanoscience. But at that time any type 
of measurement in such a small scale was a long-sought goal. Study at 
molecular scale level is not a simple one as we cannot avoid the effect 
of interface to the external electrodes. However, the progress in the 
theoretical works~\cite{nardelli} was continuing, which bestowed 
inspirations to the experimentalists to take such task as a challenge. 
Now with the advancement in nanotechnology, it is possible to investigate 
several transport properties not only through a group of 
molecules~\cite{dadosh} but also through a single molecule~\cite{reed}. 
This single molecular electronics may play a key role in designing 
nanoelectronic circuits. For this we have to have a thorough understanding 
of the electronic transport processes at this molecular scale 
level~\cite{xue1,baer1,baer2,baer3,walc1,walc2}. Many problems are 
yet to be solved to make this field much more reliable. Therefore, the 
electron transport in molecular systems is an open area and detailed 
investigations of molecular transport are still needed. 

All these works we have referred above are related to two-terminal 
electron transports. We can also analyze various transport phenomena 
of a multi-terminal system, which was first addressed by 
B\"{u}ttiker~\cite{buttiker}. The B\"{u}ttiker formalism, which is an 
extension of the Landauer two-terminal conductance formula, is a very 
simple and elegant way to divulge the transport mechanism in terms of 
various transmission probabilities. There are several pioneering 
works~\cite{xu1,stafford,sun,zhao,emberly} based on this formalism, 
which are very interesting from the theoretical as well as experimental
point of view.

Several {\em ab initio} methods~\cite{damle, derosa, taylor, xue2} are 
there which may be utilized to study electron transport properties through 
molecular junctions. At the same time, tight-binding model has been 
extended to density functional theory (DFT) for transport 
calculations~\cite{tagami1}. But in case of molecular systems, the 
investigations based on this theory (DFT) have some quantitative 
discrepancies compared to the experimental predictions. More over, these 
{\em ab initio} theories are computationally very expensive. To avoid 
this we do model calculations by using a simple tight-binding framework.

In the present article we do a theoretical study of multi-terminal 
electron transport through a single phenalenyl molecule~\cite{tagami2,
tagami3} attached to semi-infinite one-dimensional ($1$D) metallic leads. 
We do exact numerical calculations based on single particle Green's 
function formalism~\cite{san1,san2} to evaluate conductance, reflection 
probability and current-voltage characteristics. Quite interestingly, we 
show that the positions where the leads are connected to the molecule as
well as the presence of other leads have eloquent effects on these 
transport properties. More over, these characteristics are also 
influenced significantly by the molecule-to-lead coupling strengths. 
These aspects can be utilized in designing nano-electronic devices. 

We organize the paper as follows. With a brief introduction (Section 
$1$), in Section $2$ we describe our model and the theoretical 
background. Results are analyzed in Section $3$. Finally we conclude 
our results in Section $4$.

\section{Model and a view of theoretical formulation}

\begin{figure}[ht]
{\centering \resizebox*{7.3cm}{3cm}{\includegraphics{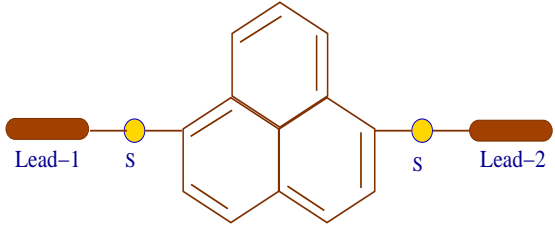}}\par}
\caption{(Color online). Two-terminal quantum system. A phenalenyl molecule 
is attached symmetrically to two semi-infinite $1$D metallic leads, viz, 
Lead-1 and Lead-2 through thiol (SH bond) groups in the chemisorption 
technique where sulfur (S) atoms reside and hydrogen (H) atoms remove. 
The filled yellow circles correspond to the location of S atoms.}
\label{two}
\end{figure}

\begin{figure}[ht]
{\centering \resizebox*{7.2cm}{5.3cm}{\includegraphics{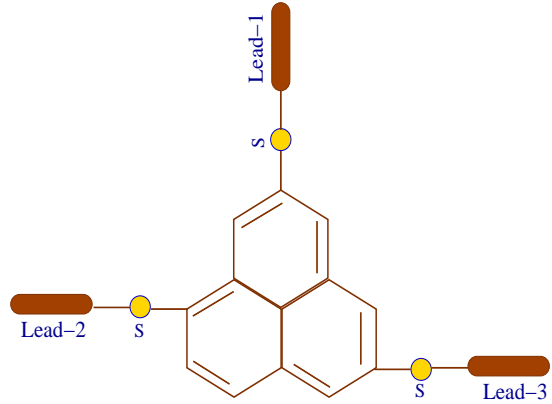}}\par}
\caption{(Color online). Three-terminal quantum system. A phenalenyl 
molecule is attached asymmetrically to three semi-infinite $1$D metallic 
leads, namely, Lead-1, Lead-2 and Lead-3 through sulfur (S) atoms.}
\label{three}
\end{figure}

\begin{figure}[ht]
{\centering \resizebox*{5.1cm}{7.8cm}{\includegraphics{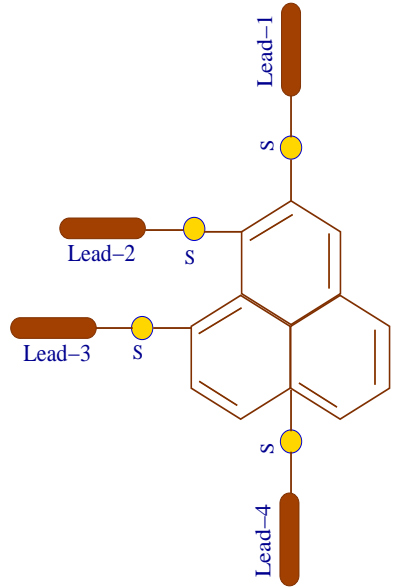}}\par}
\caption{(Color online). Four-terminal quantum system. A phenalenyl 
molecule is attached asymmetrically to four semi-infinite $1$D metallic 
leads, viz, Lead-1, Lead-2, Lead-3 and Lead-4 through sulfur (S) atoms.}
\label{four}
\end{figure}

In this section we focus our attention on the systems where a single 
phenalenyl molecule is attached symmetrically or asymmetrically to 
semi-infinite $1$D metallic leads through thiol (SH bond) groups. The 
models are shown schematically in Figs.~\ref{two}, \ref{three} and 
\ref{four} where, the number of leads attached to the molecule is $2$, 
$3$ and $4$, respectively. To evaluate the conductance ($g$) and current 
($I$) through this single molecular system we adopt the Green's function 
technique. For this, first we define the Green's function for the whole 
system as,
\begin{equation} 
G=\left(E - H \right)^{-1}
\label{green}
\end{equation}
where, $E = \epsilon +i \eta$ with $\eta$ arbitrarily very small number 
which can be set as zero in the limiting approximation. $\epsilon$ is the 
injecting electron energy. $H$ is the Hamiltonian of the entire system 
which is of infinite dimension. So, the above equation deals with the 
inversion of an infinite dimensional matrix corresponding to the system 
consisting of a finite size molecule and semi-infinite leads. However, 
the full Hamiltonian can be partitioned into sub-matrices that correspond 
to the individual sub-systems like,  
\begin{equation}
H = H_M + \sum \limits_{p=1}^N \left(H_p + H_{pM} + H_{pM}^{\dag}\right)
\end{equation}
where, $H_M$ and $H_p$ are the Hamiltonians of the molecule and lead-p,
respectively. $N$ is the number of leads to which the molecule is attached. 
$H_{pM}$ represents the coupling matrix that will be non-zero only for the 
adjacent points in the molecular system (molecule with sulfur atoms) and 
the lead-p. Here all the leads are treated on an equal footing. Within the 
non-interacting picture, the tight-binding Hamiltonian of the molecular 
system can be manifested as,
\begin{equation}
H_M = \sum_i \epsilon_i c_i^{\dag} c_i + \sum_{<ij>} t
\left(c_i^{\dag} c_j + c_j^{\dag} c_i \right)
\label{hamil}
\end{equation}
where, $\epsilon_i$ is the on-site energy, $t$ is the nearest-neighbor
hopping integral and $c_i^{\dag}$ $(c_i)$ is the creation (annihilation) 
operator of an electron at the site $i$. Each lead can be described by 
using a similar kind of tight-binding Hamiltonian, as given in 
Eq.~\ref{hamil}, characterized by two parameters $\epsilon_0$, the 
on-site potential and $t_0$, the nearest-neighbor hopping integral. 

Following the partition of the Hamiltonian, the Green's function can also 
be partitioned into sub-matrices and the effective Green's function for 
the molecular system can be indited (using Lowdin's partitioning 
technique~\cite{lowdin1,lowdin2}) as,
\begin{equation}
G_M=\left(E - H_M -\sum\limits_{p=1}^N \Sigma_p \right)^{-1}
\label{greenmolecule}
\end{equation}
where, $\Sigma_p$ is the self-energy due to the coupling of the molecular 
system to the lead-p. It is straightforward to obtain an explicit 
expression for self-energy corresponding to lead-p,
\begin{equation}
\Sigma_p = H^{\dag}_{pM} G_p H_{pM}
\label{sigma}
\end{equation}
where, $G_p=\left(E-H_p\right)^{-1}$ is the Green's function of lead-p. All 
the coupling information are inscribed into this self-energy expression.
Once the Green's function is established, the coupling function $\Gamma_p$ 
can be easily obtained from the equation~\cite{datta1,datta2}, 
\begin{equation}
\Gamma_p(E) = i \left[\Sigma^r_p(E)-\Sigma^a_p(E)\right]
\label{gamma}
\end{equation}
where, the advanced self-energy $\Sigma^a_p$ is the Hermitian conjugate of 
the retarded self-energy $\Sigma^r_p$. Thus, we can write,
\begin{equation}
\Gamma_p= -2 \mbox{Im} \left(\Sigma^r_p\right)
\label{gammasd}
\end{equation}
In order to evaluate the conductance for the multi-terminal quantum
system, we use the B\"{u}ttiker formalism~\cite{datta1}, valid
at much low temperature and bias voltage, in the form,
\begin{equation}
g_{pq} = \frac{2 e^2} {h} T_{pq} 
\label{conduc}
\end{equation}
where, $T_{pq}$ is the transmission probability of an electron across the 
molecular system from the lead-p to lead-q and it is related to the 
reflection probability by the equation,
\begin{equation}
R_{pp} + \sum \limits_{q(\neq p)}T_{qp}=1
\label{reflec}
\end{equation}
which is obtained from the condition of current conservation~\cite{xu2}. 
Now, this transmission probability can be expressed in terms of the 
effective Green's function of the molecular system and molecule-to-lead 
coupling as,
\begin{equation}
T_{pq} = {\mbox{Tr}} \left[\Gamma_p G^r_M \Gamma_q G^a_M \right]
\label{trans}
\end{equation}
where, $G^r_M$ and $G^a_M$ are the retarded and advanced Green's functions 
of the molecular system, respectively. $\Gamma_p$ and $\Gamma_q$ 
represent the couplings of the molecule to the lead-p and lead-q, 
respectively. Since the coupling matrix $H_{pM}$ is non-zero only for 
the adjacent points, $n$ and $m$, the transmission probability 
becomes~\cite{mujica},
\begin{equation}
T_{pq} = 4~ \Delta_p(nn)~ \Delta_q(mm) \mid{G_M(nm)}\mid^2
\end{equation}
where, $\Delta_p(nn)=\langle n | \Delta_p | n \rangle$, 
$\Delta_q(mm)=\langle m | \Delta_q | m \rangle$,
$G_M(nm)=\langle n | G_M | m \rangle$ and $\Delta_p$, $\Delta_q$ are 
the imaginary parts of $\Sigma_p$ and $\Sigma_q$, respectively.

In case of two-terminal system~\cite{maiti}, Eq.~\ref{conduc} becomes 
quite simpler like,
\begin{equation}
g = \frac{2 e^2} {h} T
\label{land}
\end{equation}
and accordingly, the reflection probability becomes $R=1-T$. For the
two-terminal quantum system the above expression (Eq.~\ref{land}) is the
so-called Landauer conductance formula. 

The current $I_p$ passing through the lead-p can be obtained from the 
following expression~\cite{datta1},
\begin{equation}
I_p=\frac{2 e} {h} \sum\limits_q \int\limits_{-\infty}^{\infty} T_{pq}(E)
\left [f_p(E)- f_q(E) \right]~dE
\label{current}
\end{equation}
where, $f_{p(q)}=f\left(E-\mu_{p(q)}\right)$ is the Fermi distribution 
function with the chemical potential $\mu_{p(q)}=E_F \pm e V_{p(q)}/2$. 
$E_F$ is the equilibrium Fermi energy. Throughout this calculation, we 
assume that the entire voltage is dropped across the molecule-lead 
interfaces as this assumption introduces a minimal effect on the 
behavior of the $I$-$V$ characteristics. In this article we set 
$c=h=e=1$ for the sake of simplicity.

\section{Numerical results and discussion}

In order to illustrate the results, let us first mention the values of
different parameters used in our numerical calculations. All the on-site 
energies of molecule, sulfur atom and leads are set to zero, while the 
nearest-neighbor hopping strengths are fixed at 3 for both the molecule 
($t$) and leads ($t_0$). But the values of molecule-to-lead coupling 
strengths ($\tau_p$ for lead-p) are different from the value assigned 
for $t$ and $t_0$. Based on the molecular coupling strength, we analyze 
our results in two distinct regimes. One is the weak-coupling regime and 
the other is the strong-coupling regime. In the first case, $\tau_p<<t$ 
and we set $\tau_p=0.5$. In the second case, $\tau_p \sim t$ and for this
regime we fix $\tau_p=2.5$. Here we consider that $\tau_p$'s are identical
for all the leads $p$ and set the equilibrium Fermi energy $E_F$ at
$0$.

\subsection{Conductance-energy characteristics}

In the forthcoming sub-sections we present the characteristic properties 
of electron transport for two-, three- and four-terminal molecular systems 
where the molecules are attached to leads via thiol-linking groups (SH bond). 
In experiments, leads are generally designed from gold (Au) and the thiol 
groups are linked by using chemisorption technique~\cite{holleitner} where 
hydrogen (H) atoms remove and sulfur (S) atoms survive. 

\subsubsection{Two-terminal conductance}

\begin{figure}[ht]
{\centering \resizebox*{8.2cm}{6.9cm}{\includegraphics{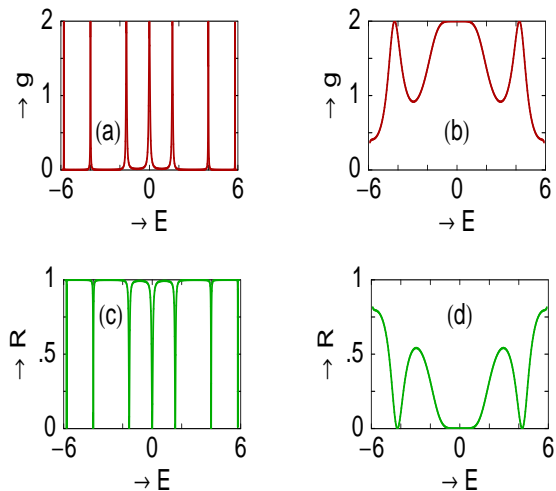}}\par}
\caption{(Color online). Conductance $g$ (red lines) and reflection 
probability $R$ (green lines) as functions of energy $E$ for two-terminal 
molecular system. (a) and (c) represent the results for the weak-coupling 
limit, while (b) and (d) correspond to the same for the strong-coupling 
limit.} 
\label{condtwo}
\end{figure}

In Fig.~\ref{condtwo} we present the variation of two-terminal conductance 
$g$ (red curves) and reflection probability $R$ (green curves) with 
injecting electron energy $E$. The results in the weak molecule-to-lead 
coupling limit are shown in (a) and (c), while (b) and (d) represent the 
same for the strong molecule-to-lead coupling limit.
In the weak-coupling regime, the presence of sharp resonant peaks indicates 
that electron transmission occurs at some typical energy values, while
for all other energies conductance vanishes (see Fig.~\ref{condtwo}(a)). 
All these resonant peaks are associated with the energy eigenvalues of 
the molecular system, and therefore, we predict that conductance spectrum 
is a fingerprint of the electronic structure of the system. Most of the 
resonant peaks approach the value $2$, the maximum value of $g$ following 
the Landauer conductance formula (see Eq.~\ref{land}) and hence $T$ 
goes to unity at these resonances indicating ballistic transmission through 
the molecular wire. But the behavior of conductance spectrum changes in 
case of the strong molecule-to-lead coupling limit. Width of each resonant 
peak becomes larger and larger as we increase gradually the molecule-to-lead 
coupling strength and for a large molecular coupling, we have the situation 
where electron transmission takes place for the entire energy range (for
illustration, see Fig.~\ref{condtwo}(b)). The effect of such broadening
comes from the imaginary parts of the self-energies~\cite{datta1}. All 
these phenomena emphasize that fine tuning in the energy scale is necessary 
as long as the coupling strength is much weak, while it is not required in 
case of the strong molecule-to-lead coupling limit.    

This scenario is just inverted in case of reflection probability $R$. 
In the weak-coupling regime, sharp dips appear (Fig.~\ref{condtwo}(c)) for 
some particular energy values where conductance shows resonant peaks, as 
$R$ follows the simple relation $R=1-T$ for the two-terminal molecular 
system. For all other energy values, $R=1$ indicating no transmission of 
electron across the molecule. The effect of molecule-to-lead coupling 
is exactly similar to that for conductance spectrum. In the strong-coupling
regime (see Fig.~\ref{condtwo}(d)), the reflection probability no longer 
reach the maximum value ($1$) for the entire energy range. 

\subsubsection{Three-terminal conductance}

\begin{figure}[ht]
{\centering \resizebox*{8.3cm}{10.8cm}{\includegraphics{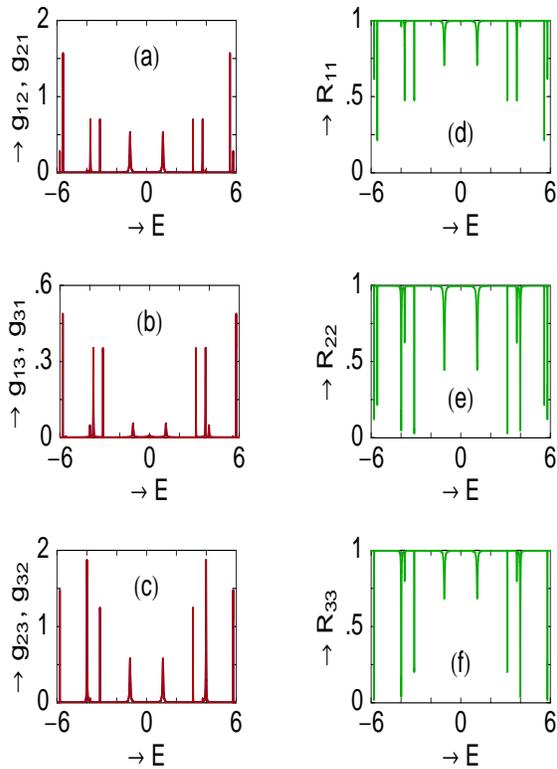}}\par}
\caption{(Color online). Conductance $g_{pq}$ (red curves) and reflection 
probability $R_{pp}$ (green curves) as functions of energy $E$ for 
three-terminal molecular system. All the results are presented only
for the weak-coupling regime.}
\label{condthree}
\end{figure}

To illustrate the results for the three-terminal quantum system, constructed 
by attaching three leads to the molecule (see Fig.~\ref{three}), let us 
start by referring to Fig.~\ref{condthree} where the first column shows 
the conductance spectra $g_{pq}$ (from lead-p to lead-q) and the second 
column presents the nature of reflection probability $R_{pp}$. Similar 
to the two-terminal molecular system, some sharp peaks appear in the 
conductance spectra. But the point is that for the three-terminal system
most of the resonant peaks do not reach the value $2$. From the conductance 
spectra it is clear that the heights are much reduced compared to the 
two-terminal case. This is solely due to the effect of quantum 
interference of the electronic waves passing through different arms of
the molecular rings. In this three-terminal system, three leads are 
attached asymmetrically to the molecule at three different 
locations, which provide different path ways for electron transmission
between the leads. This introduces anomalous features in the conductance 
spectra as illustrated in Figs.~\ref{condthree}(a), (b) and (c). More 
over, in this three-terminal case, nature of variations of reflection 
probabilities is not so simple as we get in the case of two-terminal 
molecular system. Here, it 
is not necessary that $R_{pp}$ shows dips or peaks where $g_{pq}$ has 
peaks or dips since it ($R_{pp}$) depends on the combined effect of 
$T_{pq}$'s obeying the expression given in Eq.~\ref{reflec}. Another 
important point we like to mention here is that, one can easily find 
$T_{qp}$ for any two leads $q$ and $p$ if $T_{pq}$ is known since
the relation $T_{pq}=T_{qp}$ holds true following the time-reversal 
symmetry. The effect of molecule-to-lead coupling strength is 
identical to the case of two-terminal conductance and therefore, we do 
not show those results further.

\subsubsection{Four-terminal conductance}

The transport properties of the four-terminal molecular system is also 
\begin{figure}[ht]
{\centering \resizebox*{8.3cm}{11cm}{\includegraphics{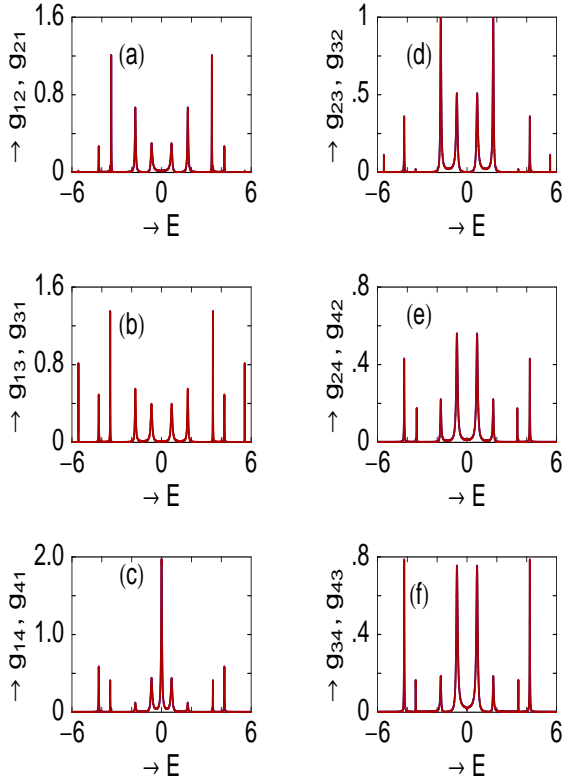}}\par}
\caption{(Color online). Four-terminal conductance $g_{pq}$ as functions 
of energy $E$ in the limit of weak molecular coupling.}
\label{condfour}
\end{figure}
described by investigating 
various conductances $g_{pq}$ and reflection probabilities $R_{pp}$, which 
are obtained from the Eqs.~\ref{conduc} and~\ref{reflec}, respectively. 
The results are plotted in Fig.~\ref{condfour} considering the weak 
molecule-to-lead coupling. In each figure $g_{pq}$ and $g_{qp}$ are 
plotted and they are superposed to each other due to their symmetry. 
Sharp resonant peaks of different heights for some particular energies
appear in the conductance spectra similar to the two- or three-terminal 
conductance spectra. From the conductance spectra, the effect of quantum 
interference associated with the molecule-to-lead interface geometry is
well understood. 
\begin{figure}[ht]
{\centering \resizebox*{8.3cm}{8cm}{\includegraphics{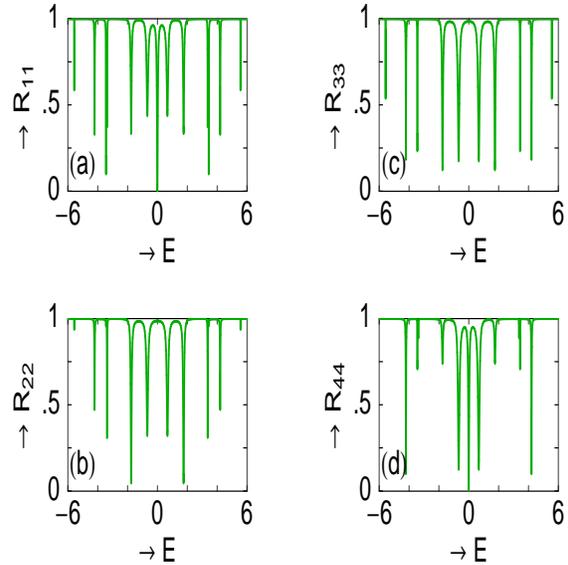}}\par}
\caption{(Color online). Reflection probability $R_{pp}$ as a function of 
energy $E$ for four-terminal quantum system. These results correspond to 
the weak-coupling regime.} 
\label{reflection}
\end{figure}
Also the dependence of electron transport on molecular coupling strength 
is exactly similar to that in Fig.~\ref{condtwo} and accordingly, the
results are not given further. Comparing the results given in 
Figs.~\ref{condthree} and \ref{condfour} we observe that, the conductances
for two particular leads located at the same positions of the molecule 
exhibit completely different features in presence of the other leads. For 
instance, $g_{13}$ of the four-terminal system and $g_{12}$ of the 
three-terminal system show different spectra though these two conductances
are evaluated for the two leads located at the same positions of the
molecule.  
Similarly, careful observation depicts that pathways between the lead-$2$ 
and lead-$3$ of the three-terminal system are exactly similar to that 
between lead-$2$ and lead-$4$ of the four-terminal system. In spite of 
this the corresponding spectra i.e., $g_{23}$ of Fig.~\ref{condthree} 
and $g_{24}$ of Fig.~\ref{condfour} are different from each other due to 
the presence of the fourth lead. The effect of the leads are incorporated 
into the self-energies which lead to the change in conductance spectra.
Thus, we can say that the presence of additional leads has a pronounced 
effect on the electron transport properties. In $R$-$E$ spectra given
in Fig.~\ref{reflection}, we get various dips at different energies 
depending on the combined effect of the transmission probabilities 
(Eq.~\ref{reflec}), similar to the case of three-terminal system. 
All the reflection probabilities are calculated only for the limit
of weak molecular coupling. Exactly similar feature, except the
broadening, will be observed in the case of strong-coupling.

\subsection{Current-voltage characteristics}

The scenario of electron transport through molecular junction becomes 
much more transparent when we discuss the current-voltage ($I$-$V$) 
characteristics, where the current is evaluated by integrating the 
transmission probability $T$ using Eq.~\ref{current}. The nature of 
the variation of transmission probability is exactly similar to that 
of conductance spectra except the factor $2$ as we have assumed $e=h=1$ 
in the Landauer conductance formula (Eq.~\ref{land}). Here, we discuss 
the $I$-$V$ characteristics for two-, three- and four-terminal molecular 
systems separately in the following sub-sections. 

\subsubsection{Two-terminal molecular system}

\begin{figure}[ht]
{\centering \resizebox*{7.8cm}{10cm}{\includegraphics{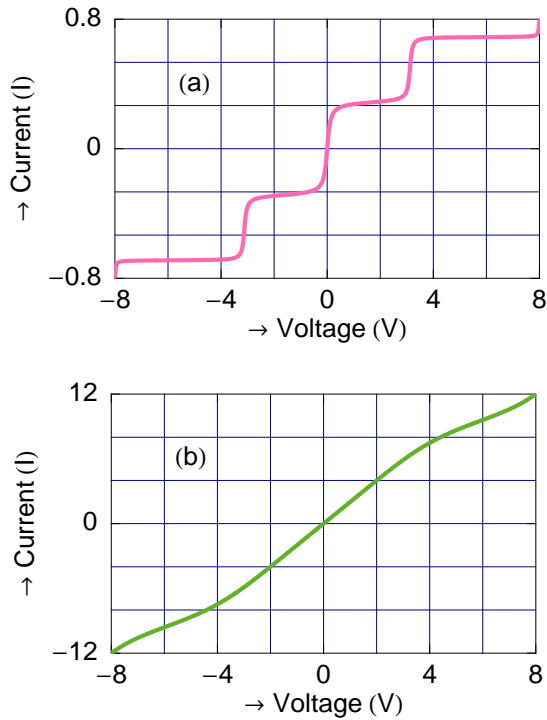}}\par}
\caption{(Color online). Current $I$ as a function of applied bias voltage 
$V$ for two-terminal molecular system. (a) Weak-coupling limit and (b) 
strong-coupling limit.}
\label{currtwo}
\end{figure}

As an illustration, we display the $I$-$V$ characteristics for the 
two-terminal system in Fig.~\ref{currtwo}, where (a) and (b) correspond 
to the results for the weak and strong molecule-to-lead coupling limits, 
respectively. For this two-terminal case, current can be expressed 
mathematically as follows,
\begin{eqnarray}
I &=& g~ (V_1-V_2) \nonumber \\
&=& g~ V_{12}
\end{eqnarray}
where, $V_{12}$ is the voltage difference between the lead-1 and lead-2.
In the case of two-terminal molecular system, we have attached two leads 
to the molecule and their chemical potentials are changed as the bias 
voltage is applied. With the increase of the voltage, the gap increases 
more and more and eventually crosses molecular energy levels 
one after another. Accordingly, current channels are opened up and jumps 
in the $I$-$V$ curve appear. This provides staircase-like structure in
the current-voltage spectrum and only for the case of weak molecular 
coupling these sharp steps appear. But this feature i.e., step-like changes 
gradually towards continuous nature with the increase of molecule-to-lead 
coupling strength. In addition to that, current amplitude becomes much 
higher compared to the case of weak-coupling limit. By noting the area 
under the curve of Fig.~\ref{condtwo}(b) the reason behind this enhancement 
of current amplitude is clearly understood. Thus, it can be manifested that 
molecule-to-lead coupling strength has a significant influence on molecular
transport. 

\subsubsection{Three-terminal molecular system}

Now, we describe the current-voltage characteristics for the three-terminal 
system where we find the current $I_p$ in lead-p by integrating the 
\begin{figure}[ht]
{\centering \resizebox*{7.4cm}{4.75cm}{\includegraphics{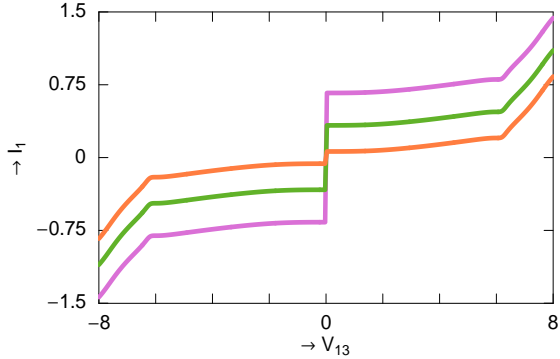}}\par}
\caption{(Color online). $I_1$ as a function of $V_{13}$ ($=V_1-V_3$) 
for the three-terminal molecular system in the limit of strong-coupling, 
keeping $V_{12}$ ($=V_1-V_2$) as a constant. The orange, green and magenta 
colors correspond to $V_{12}=1$, $2$ and $3$, respectively.}
\label{3curri1}
\end{figure}
transmission function $T_{pq}$ considering the effects of other terminals. 
To be more precise, we write the current expressions for three different
leads as follows,
\begin{eqnarray}
I_1 & = & g_{12} \left(V_1-V_2\right)+g_{13}\left(V_1-V_3\right)\nonumber\\
 & = & g_{12} V_{12}+g_{13} V_{13} \label{threecurr1} \\
I_2 & = & g_{21} \left(V_2-V_1\right)+g_{23}\left(V_2-V_3\right)\nonumber\\
 & = & g_{21} V_{21}+g_{23} V_{23} \label{threecurr2} \\
I_3 & = & g_{31} \left(V_3-V_1\right)+g_{32}\left(V_3-V_2\right)\nonumber\\
 & = & g_{31} V_{31}+g_{32} V_{32}
\label{threecurr3}
\end{eqnarray}
where, $V_{pq}=\left(V_p-V_q\right)$ is the voltage difference between the 
two leads named as lead-p and lead-q.

In Fig.~\ref{3curri1}, we plot $I_{1}$ for the lead-1 as a function of 
$V_{13}$ for different fixed values of $V_{12}$ in the strong 
molecule-to-lead coupling limit. The orange, green and magenta curves 
represent the currents for $V_{12}=1$, $2$ and $3$, respectively. It is 
clear from the figure that for a constant value of $V_{12}$, the moment 
we switch on the bias voltage between lead-1 and lead-3, current rises to 
a large value. Then, for a wide range of $V_{13}$, it ($I_1$) slowly 
increases with the rise of $V_{13}$ and finally the rate of increment 
\begin{figure}[ht]
{\centering \resizebox*{7.4cm}{4.75cm}{\includegraphics{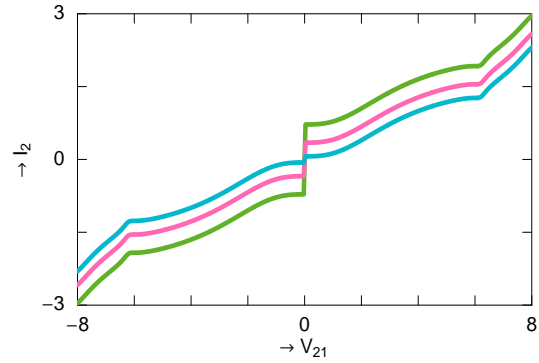}}\par}
\caption{(Color online). $I_2$ as a function of $V_{21}$ ($=V_2-V_1$) for
the three-terminal molecular system in the strong-coupling regime, 
keeping $V_{23}$ ($=V_2-V_3$) as a constant. The blue, pink and green 
lines correspond to $V_{23}=1$, $2$ and $3$, respectively.}
\label{3curri2}
\end{figure}
of the current gets enhanced when $V_{13}$ is quite high. This behavior
i.e., the rate of increment of the current with bias voltage solely depends
on the positions of resonant peaks in the $g$-$E$ spectrum. For a particular
value of $V_{13}$, $I_1$ increases as we increase $V_{12}$ which is clearly 
visible from three different curves plotted in Fig.~\ref{3curri1}. 

The variation of current $I_2$ through lead-2 as a function of $V_{21}$ 
is shown in Fig.~\ref{3curri2}, keeping the voltage $V_{23}$ as a constant. 
The currents are evaluated in the strong-coupling limit, where the blue, 
pink and green lines correspond to $V_{23}=1$, $2$ and $3$, respectively.

\begin{figure}[ht]
{\centering \resizebox*{7.4cm}{4.75cm}{\includegraphics{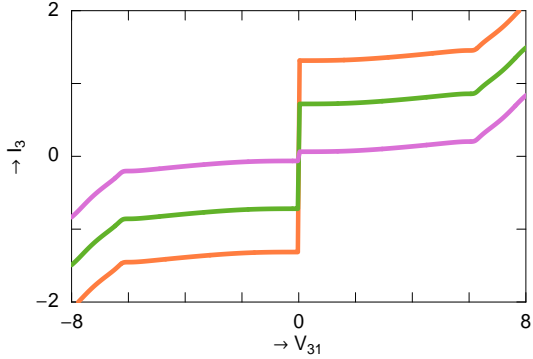}}\par}
\caption{(Color online). $I_3$ as a function of $V_{31}$ ($=V_3-V_1$) for
the three-terminal molecular system in the strong molecule-to-lead 
coupling limit, keeping $V_{32}$ as a constant. The magenta, green and 
orange colors correspond to $V_{32}=1$, $3$ and $5$, respectively.}
\label{3curri3}
\end{figure}

In a similar fashion in Fig.~\ref{3curri3} we display $I_3$-$V_{31}$
characteristics for different fixed values of $V_{32}$ considering the
case of strong-coupling limit. The magenta, green and orange curves
correspond to $V_{32}=1$, $3$ and $5$, respectively. 

The characteristic features of the currents $I_1$, $I_2$ and $I_3$ passing 
through three different leads are quite analogous to each other. Depending
on the conductance-energy spectra, we get different current amplitudes for
three different leads which are clearly observed from the results presented
in Figs.~\ref{3curri1}, \ref{3curri2} and \ref{3curri3}. All these currents
are computed only for the strong-coupling limit. We can also determine
the currents for the limit of weak-coupling and in that case we will get
sharp step-like features as a function of bias voltage with much reduced 
amplitude compared to the strong-coupling case. The origin of step-like
behavior in current is clearly mentioned in the case of two-terminal 
molecular system (Sec. 3.2.1).

From the above current expressions (Eqs.~\ref{threecurr1}, \ref{threecurr2}
and \ref{threecurr3}) we see that the current in anyone lead is related
to two potential functions. For instance in Eq.~\ref{threecurr1}, there
are two parameters like $V_{12}$ and $V_{13}$. Keeping $V_{12}$ as a 
constant we plot the current $I_1$ in terms of $V_{13}$ (see 
Fig.~\ref{3curri1}). At the same time, we can also draw the $I_1$-$V_{12}$
characteristics, considering $V_{13}$ as a constant. Both for these two 
cases, the characteristic features are quite similar. This argument is
also valid for the other two current expressions (Eqs.~\ref{threecurr2}
and \ref{threecurr3}).

All the above features of current-voltage characteristics in this 
three-terminal molecular system clearly support the basic features
of a traditional macroscopic transistor. Thus we can predict that 
the three-terminal molecular system may be utilized to design a 
nano-scale molecular transistor.

\subsubsection{Four-terminal molecular system}

Finally, we focus our attention on the current-voltage characteristics for
\begin{figure}[ht]
{\centering \resizebox*{7.4cm}{4.75cm}{\includegraphics{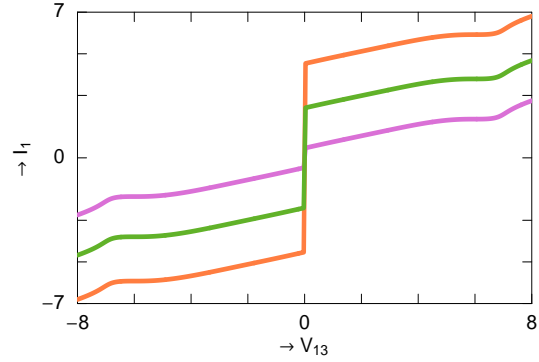}}\par}
\caption{(Color online). $I_1$ as a function of $V_{13}$ for the 
four-terminal molecular system in the strong-coupling limit, keeping
$V_{12}$ and $V_{14}$ as constant. The magenta, green and orange colors 
correspond to $V_{12}=V_{14}=1$, $3$ and $5$, respectively.}
\label{4curri1}
\end{figure}
the four-terminal molecular system. In this molecular system, the current
expressions for the four different leads are as follows,
\begin{eqnarray}
I_1 & = & g_{12} \left(V_1-V_2\right)+g_{13}\left(V_1-V_3\right)+
g_{14}\left(V_1-V_4\right) \nonumber\\
 & = & g_{12} V_{12}+g_{13} V_{13}+g_{14} V_{14} \label{fourcurr1} \\
I_2 & = & g_{21} \left(V_2-V_1\right)+g_{23}\left(V_2-V_3\right)+
g_{24}\left(V_2-V_4\right) \nonumber\\
 & = & g_{21} V_{21}+g_{23} V_{23}+g_{24} V_{24} \label{fourcurr2} \\
I_3 & = & g_{31} \left(V_3-V_1\right)+g_{32}\left(V_3-V_2\right)+
g_{34}\left(V_3-V_4\right) \nonumber\\
 & = & g_{31} V_{31}+g_{32} V_{32}+g_{34} V_{34} \label{fourcurr3} \\
I_4 & = & g_{41} \left(V_4-V_1\right)+g_{42}\left(V_4-V_2\right)+
g_{43}\left(V_4-V_3\right) \nonumber\\
 & = & g_{41} V_{41}+g_{42} V_{42}+g_{43} V_{43} 
\label{fourcurr4}
\end{eqnarray}
where, $V_{pq}$ is the voltage difference between the lead-p and lead-q.

As representative examples, in Fig.~\ref{4curri1} we plot the current in 
lead-1 ($I_1$) as a function of $V_{13}$ keeping $V_{12}$ and $V_{14}$ as
constant. The results are computed for the strong-coupling limit, where
\begin{figure}[ht]
{\centering \resizebox*{7.4cm}{4.75cm}{\includegraphics{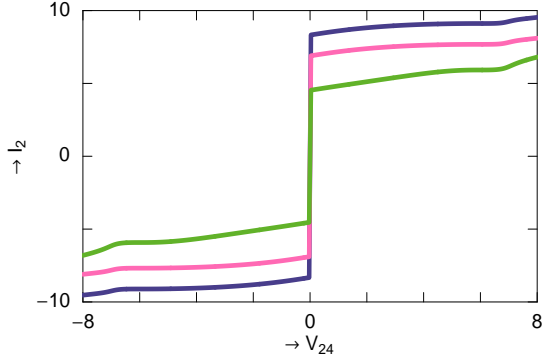}}\par}
\caption{(Color online). $I_2$ as a function of $V_{24}$ ($=V_2-V_4$) for
the four-terminal molecular system in the limit of strong-coupling, keeping
$V_{21}$ and $V_{23}$ as constant. The green, pink and dark-blue lines 
correspond to $V_{21}=V_{23}=2$, $4$ and $6$, respectively.}
\label{4curri2}
\end{figure}
the magenta, green and orange curves correspond to $V_{12}=V_{14}=1$, $3$ 
and $5$, respectively.
The variation of current $I_1$ as a function of $V_{13}$ in this
four-terminal molecular system is quite similar to that as presented in 
\begin{figure}[ht]
{\centering \resizebox*{7.4cm}{4.75cm}{\includegraphics{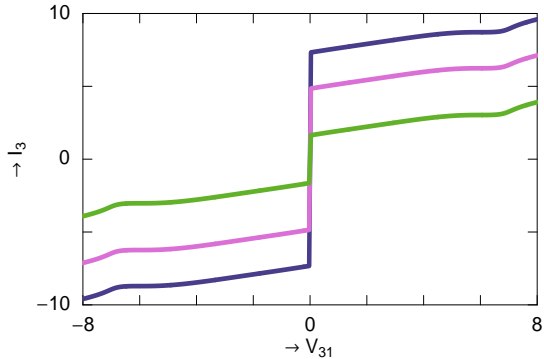}}\par}
\caption{(Color online). $I_3$ as a function of $V_{31}$ for the 
four-terminal molecular system for the case of strong-coupling limit,
considering $V_{32}$ and $V_{34}$ as constant. The green, magenta and 
dark-blue curves correspond to $V_{32}=V_{34}=1$, $3$ and $5$, respectively.}
\label{4curri3}
\end{figure}
the case of three-terminal system (Fig.~\ref{3curri1}). For a particular
value of $V_{13}$, here also the current amplitude gets increased with 
the rise of $V_{12}$ and $V_{14}$. But quite significantly we observe that,
for a particular value of $V_{13}$, the current $I_1$ passing through the
lead-1 of four-terminal system acquires much higher amplitude compared to 
the three-terminal system (see Figs.~\ref{3curri1} and \ref{4curri1}). The 
reason behind this enhancement of current amplitude is explained as follows.
From Eq.~\ref{fourcurr1} we see that $I_1$ contains three additive terms
where the contributions come from other leads, while in Eq.~\ref{threecurr1} 
there are two additive terms. The additional term appears in 
Eq.~\ref{fourcurr1} is due to the presence of fourth terminal which is
responsible for the larger current in four-terminal system compared to 
the three-terminal one.

In case of the current $I_2$ through lead-2 we show the variation with 
respect to $V_{24}$, keeping $V_{21}$ and $V_{23}$ fixed to a particular 
\begin{figure}[ht]
{\centering \resizebox*{7.4cm}{4.75cm}{\includegraphics{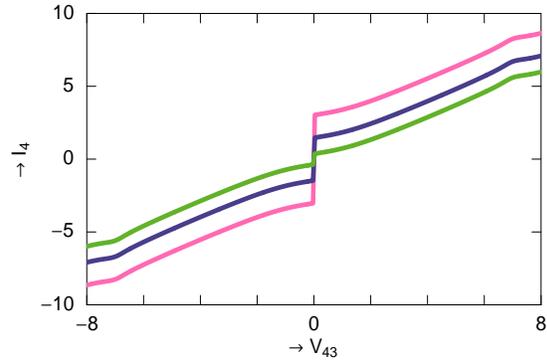}}\par}
\caption{(Color online). $I_4$ as a function of $V_{43}$ ($=V_4-V_3$) for
the four-terminal molecular system, keeping $V_{41}$ and $V_{42}$ as 
constant. The currents are evaluated in the strong-coupling limit,
where the green, dark-blue and pink lines correspond to $V_{41}=V_{42}=1$, 
$5$ and $9$, respectively.}
\label{4curri4}
\end{figure}
value as presented in Fig.~\ref{4curri2}. The currents are determined in 
the strong-coupling regime, where the green, pink and dark-blue lines 
correspond to $V_{21}=V_{23}=2$, $4$ and $6$, respectively. 

Similarly, in Fig.~\ref{4curri3} we display $I_3$-$V_{31}$ characteristics
considering $V_{32}$ and $V_{34}$ as constants in the limit of 
strong-coupling. The green, magenta and dark-blue lines represent the
currents for $V_{32}=V_{34}=1$, $3$ and $5$, respectively.

At the end, in Fig.~\ref{4curri4} we show the variation of current $I_4$ 
passing through lead-4 as a function of $V_{43}$ in the limit of strong
molecular coupling, when $V_{41}$ and $V_{42}$ are kept as constants. The 
green, dark-blue and pink curves correspond to the currents for 
$V_{41}=V_{42}=1$, $5$ and $9$, respectively.
 
Similar to the case of three-terminal molecular system, for this 
four-terminal case, we can also plot the current through any lead-p in 
aspect of the other potential functions as given in Eqs.~\ref{fourcurr1}, 
\ref{fourcurr2}, \ref{fourcurr3} and \ref{fourcurr4}. For all these 
cases, the characteristic features are very much similar those are
presented in Figs.~\ref{4curri1}, \ref{4curri2}, \ref{4curri3} and 
\ref{4curri4}, and hence, we do not re-plot the results further.

\section{Closing remarks}

In the present paper we have used a parametric approach to study
multi-terminal electron transport through a single phenalenyl molecule.
Using a simple tight-binding framework we have performed all the
numerical calculations through single particle Green's function 
formalism. The basic features of electron transport in this molecular
system are explored by investigating the multi-terminal conductance, 
reflection probability and current. Following a detailed description
of electron transport in two-terminal quantum system, we have revealed
the essential features of electron transport in the three- and 
four-terminal quantum systems separately.

Our clear investigation predicts that the electron transport in
multi-terminal molecular system significantly depends on (a) the 
molecule-lead interface geometry, (b) the presence of other leads and 
(c) the strength of molecular coupling to the side attached leads. The 
unique characteristics of this phenalenyl molecule with a very small 
size has enhanced the importance of the present article. Our parametric 
study provides several significant features to reveal electron transport 
through any complicated multi-terminal quantum system.
  
In the present paper we have done all the calculations by ignoring
the effects of temperature, electron-electron correlation, etc. We need 
further study by incorporating all these effects.

%%%%%%%%%%%%%%%%%%%%%%%%%%%%%

\end{document}